\documentclass[10pt,conference]{IEEEtran}
\usepackage{times}
\usepackage{graphicx,url}
\usepackage{color}
\usepackage{latexsym}

\usepackage[utf8]{inputenc}

\usepackage{epsfig}
\usepackage{subfigure}
\usepackage{amssymb}
\usepackage{amsmath}
\usepackage{multicol}
\usepackage{indentfirst}
\usepackage{float}
\usepackage{algorithm,algorithmic}

\title{Cross-Pollination of Information in Online Social Media: A Case Study on Popular Social Networks}
\author{Paridhi Jain$^{\**}$, Tiago Rodrigues$^{\dagger}$, Gabriel Magno$^{\dagger}$, Ponnurangam Kumaraguru$^{\**}$, Virgílio Almeida$^{\dagger}$  \newline
\\ { $^{\**}$ Indraprastha Institute of Information Technology, Delhi (IIIT-D), India} \\ $^{\dagger}$ Universidade Federal de Minas Gerais (UFMG), Brazil \\ \{paridhij, pk\}@iiitd.ac.in \\ \{tiagorm, magno, virgilio\}@dcc.ufmg.br}
 
\begin{document}

\maketitle

%\keywords{twitter, information propagation, social network}

%\category{J.4.} {Computer Applications} Social and behavioral sciences {Miscellaneous}
%\category{H.3.5}{Online Information Services}{Web-based services}
%\terms{Human factors, Measurement}
%\keywords{Twitter, word-of-mouth, information propagation, social networks}

%\input{abstract} % abstract
\begin{abstract}
\noindent Owing to the popularity of Online Social Media (OSM), Internet users share a lot of information (including personal) on and across OSM services every day. For example, it is common to find a YouTube video embedded in a blog post with an option to share the link on Facebook. Users recommend, comment, and forward information they receive from friends, contributing in spreading the information in and across OSM services. We term this information diffusion process from one OSM service to another as \emph{Cross-Pollination}, and the network formed by users who participate in Cross-Pollination and content produced in the network as \emph{Cross-Pollinated network}. Research has been done about information diffusion within one OSM service, but little is known about Cross-Pollination. In this paper, we aim at filling this gap by studying how information (video, photo, location) from three popular OSM services (YouTube, Flickr and Foursquare) diffuses on Twitter, the most popular microblogging service. Our results show that Cross-Pollinated networks follow temporal and topological characteristics of the diffusion OSM (Twitter in our study). Furthermore, popularity of information on source OSM (YouTube, Flickr and Foursquare) does not imply its popularity on Twitter. Our results also show that Cross-Pollination helps Twitter in terms of traffic generation and user involvement, but only a small fraction of videos and photos gain a significant number of views from Twitter. We believe this is the first research work which explicitly characterizes the diffusion of information across different OSM services.
\end{abstract}

%\vspace{-5pt}
%\input{introduction} % introduction
\section{Introduction} \label{sec:introduction}

\noindent Online Social Media (OSM) has become a popular medium to share information, gaining an explosive growth in terms of user base and content. For example, Twitter user base increased by 1,444\% in 2009, and more than 100 million new users signed up in 2010~\cite{twolyno,htucm}. Users spend 700 million minutes / month on Facebook~\cite{facebook}. More than 100 hours of videos are uploaded in every 4 minutes on YouTube~\cite{smstm}. By discussing different topics every day, users form an information sharing network with online friends, who share similar interests or belong to similar domain (place, workgroup, etc). Properties of the information sharing network, increased content creation, and user participation have attracted researchers from different discipline to study OSM.

On OSM services, users create and share information with others, in a mechanism termed as \emph{information diffusion}. With users having accounts in different OSM services (e.g. YouTube, Facebook), there is a tendency to exchange information across OSM services~\cite{broxton-2010}. Finding a blog post with a Flickr photo embedded and a hyperlink to a news article as Facebook status are prominent today. Users usually post URLs on Twitter and Facebook to announce to their friends about a new blog post or a new uploaded video (on YouTube). The information diffusion process across OSM services is analogous to a process in biology, termed as \emph{Cross-Pollination}. In this process, pollen is delivered to a flower from a different plant, with the plants being different in their genesis~\cite{Darwin}. Following the same analogy, we term the information diffusion process across OSM services as \emph{Cross-Pollination}. A unit of information is analogous to pollen, and different OSM services are analogous to plants having different genesis.

Studying the dynamics and characteristics of a Cross-Pollination process is important for various reasons. Understanding Cross-Pollination can facilitate marketers to explore the rich environment for advertisement purposes. It can help social media providers to improve their systems and develop tools to facilitate the information exchange across networks. Literature about information diffusion within one OSM service can be found, but little is known about the process of exchange of information across OSM services. Several important questions are unanswered -- (1) What are the characteristics of the Cross-Pollination? (2) Does Cross-Pollination across OSM services help to increase the audience reached by the information diffused? (3) What is the relationship between the OSM services involved and how does it affect the information diffusion process? 

In this paper, we study Cross-Pollination of three popular OSM services as \emph{source OSM} -- YouTube, the largest video sharing repository; Flickr, one of the largest photo sharing repository; and Foursquare, a popular location-based social networking service -- with one another popular OSM service as \emph{diffusion OSM}, Twitter, the largest microblogging service in the world.

We define a basic unit of information as a \emph{meme}.~\footnote{A meme is an element of a culture or system of behavior passed from one individual to another by imitation or other non-genetic means, taken from \emph{http://oxforddictionaries.com/}}  A video on YouTube and a tweet on Twitter are the examples of memes. Memes can be divided into two categories: \emph{foreign} and \emph{local}. We consider all posted URLs embedding meme belonging to another OSM service as a \emph{foreign meme}. URLs embedding YouTube videos or Flickr photos, when shared on Twitter, are examples of foreign memes. We consider all other types of memes generated and diffused within one OSM service as a \emph{local meme}. Hashtags (term starting with \# to represent the topic of the tweet, e.g., \#BestDad) and mentions (internal link to another user in the form of @username) are examples of local memes on Twitter. OSM service where a foreign meme originates is termed as \emph{source OSM} (Flickr, YouTube, and Foursquare in our study), and the OSM service in which the foreign meme diffuses is termed as \emph{diffusion OSM} (Twitter in our study). A network formed by users who participate in Cross-Pollination and content produced in the network is termed as \emph{Cross-Pollinated network}.

\begin{figure}[!htb]
\includegraphics[width=.49\textwidth]{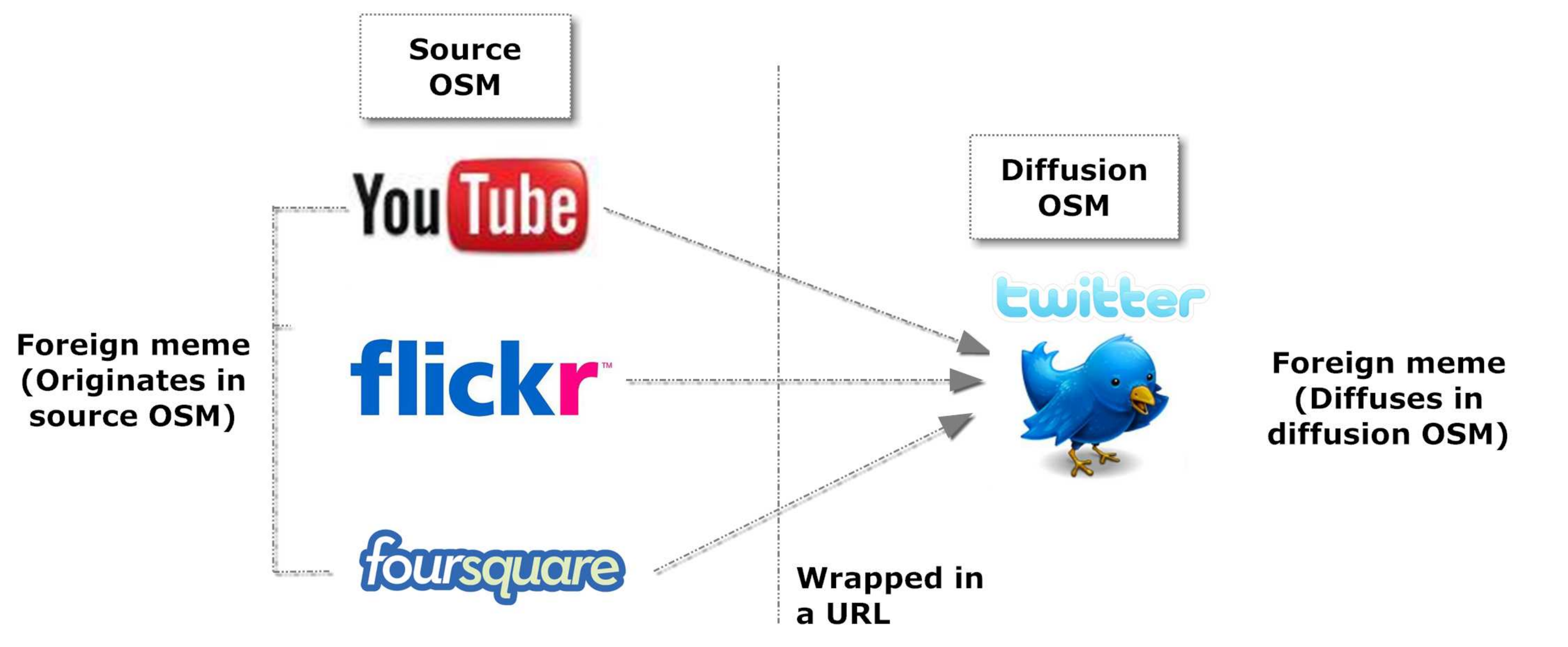}
\caption{\bf{Cross-Pollination dynamics. Shows how a foreign meme is created and diffused in a Cross-Pollinated network.}}
\label{fig:model}
\end{figure}

Twitter is a good medium to study the diffusion of foreign memes as it provides mechanisms that enable fast spreading of information. Figure~\ref{fig:model} illustrates the dynamics of exchange of memes from one OSM service to another one. Users create a meme in a source OSM, embed the meme in a URL and diffuse it on the diffusion OSM.

To the best of our knowledge, this is the first work to explicitly study this important and unexplored area of Cross-Pollination. Our main results and contributions are:
\begin{itemize}
    \item Cross-Pollinated networks follow temporal and topological characteristics of the diffusion OSM.
    \item Cross-Pollination helps only a small percentage of foreign memes to reach to large audience.
    \item Popularity of meme on source OSM does not imply its popularity on diffusion OSM and vice versa.
\end{itemize}

In the next section, we explain our data collection and methodology; in Section~\ref{sec:results}, we present the analysis and results of our study. We then present the related work in Section~\ref{sec:relatedwork}. Finally, in Section~\ref{sec:conclusion}, we conclude the paper with discussing the implications of our results, future work, and limitations of our research.

%\vspace{-3pt}
%\input{crosspollination} % define cross pollination
%\input{methodology} % methodology
\section{Methodology} \label{sec:methodology}

\noindent In this section, we describe our data collection framework and provide descriptive characteristics of the datasets used.

\subsection{Data Collection}\label{subsec:datacollection}

\begin{figure}[!ht] \begin{center}
\includegraphics[height=4cm,width=.46\textwidth]{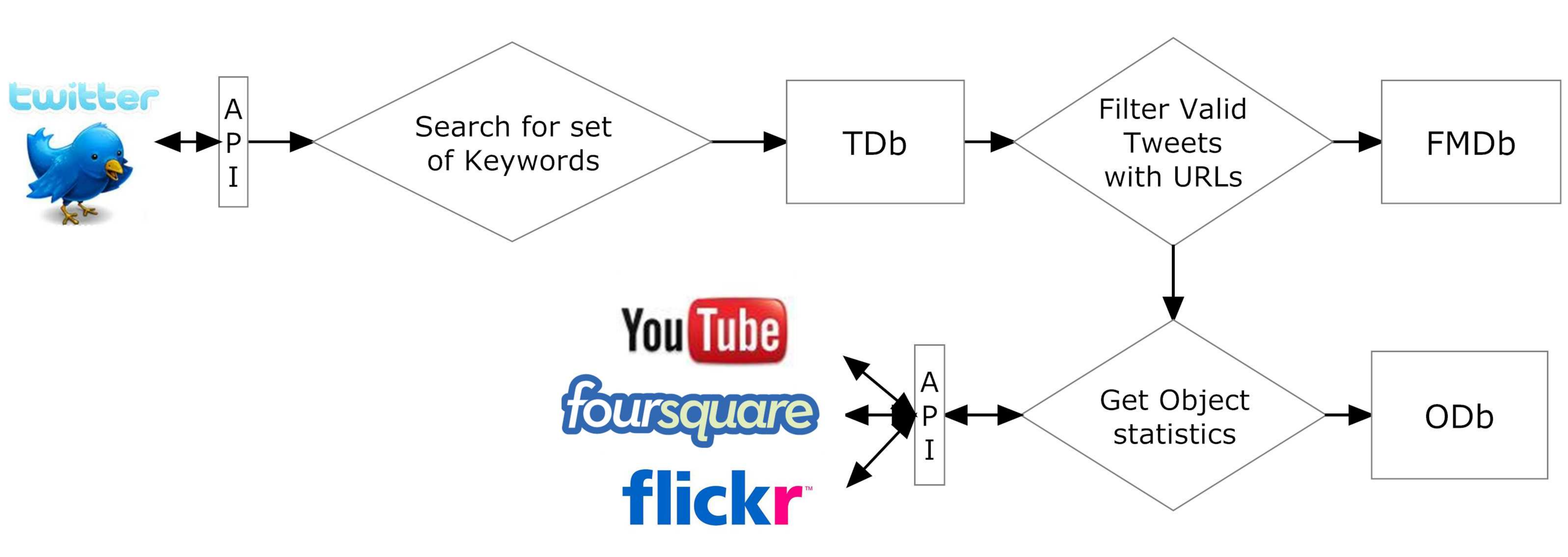}
\caption{\bf{Data collection framework. Shows the two phases of the data collection framework which monitors the events to collect data.}}
\label{fig:crawler}
\end{center}
\end{figure}

\noindent Our data collection framework is composed of two phases (see Figure~\ref{fig:crawler}). In the first phase, we used Twitter Streaming Application Program Interface (API)~\cite{twitterapi} to collect all tweets periodically, using a set of keywords. This step was part of a research project, developed by a Brazilian Research Institute,~\footnote{\emph{Instituto Nacional de Ciência e Tecnologia para a Web},  \emph{http://www.inweb.org.br/}} which tracks information about important events in several social and traditional media sources, like newspapers, blogs, and online social networks.~\footnote{The \emph{Observatório da Web} Project, \emph{http://observatorio.inweb.org.br/}} After this step, we filtered all URLs that appear on the content of the tweets.  Due to the usage of URL shorteners like \textit{http://bit.ly/},~\cite{shorteners} we expanded all shortened URLs and filtered all tweets with YouTube videos URLs, Flickr photos URLs, and Foursquare location URLs. We inserted all tweets that contain these types of URLs into \emph{Foreign meme Database}~(FMDb). In the second phase, we used YouTube~\cite{youtubeapi}, Flickr~\cite{flickrapi}, and Foursquare~\cite{foursquareapi} APIs to collect information about the foreign memes and their uploaders, storing the same in \emph{Objects Database}~(ODb).

Out of the most discussed topics on Twitter in 2010~\cite{trends2010}, we created a dataset for FIFA World Cup (FWC), a global event. The FWC is an international football competition contested by the senior men's national teams of the members of Fédération Internationale de Football Association (FIFA), the sport's global governing body. The event happens every 4 years and in 2010 it took place in South Africa, from June $11^{th}$ to July $11^{th}$. We monitored the FWC event from June $10^{th}$ to July $12^{th}$, using 112 keywords (e.g. worldcup, FIFA and southafrica) in 7 different languages (like Portuguese, English and Spanish). To ensure no data loss, we used several redundant machines to collect the same data.

\subsection{Datasets}\label{subsec:datasets}

\noindent Table~\ref{tab:datasets} presents the descriptive statistics of our datasets.  A total of 34,306 unique videos URLs were shared on Twitter during the FWC, in a total of 141,118 tweets, posted by 88,231 users. The videos were uploaded by 26,026 YouTube users. The maximum number of videos were shared on July $3^{rd}$ (3,049 videos). On this day, there were 2 important matches -- Argentina versus Germany, and Spain (future champion) versus Paraguay. On average, 1,554 videos were shared per day during the event. Table~\ref{tab:datasets} also presents statistics about Foursquare and Flickr datasets, which are less popular than YouTube on Twitter (in our dataset), but still have a representative number of URLs to study. We note that locations are not associated to owners on Foursquare, which differs from YouTube and Flickr. We also created a \emph{baseline} dataset, which contains local memes only. In total, the baseline dataset has more than 29 million tweets, created by 3.5 million users. The baseline dataset is used in several of our analysis to contrast the characteristics of Cross-Pollinated networks with the characteristics of Twitter itself. This comparison helps in understanding how the introduction of foreign memes affects the diffusion OSM.

\begin{table}[h!]
\centering
\scriptsize
\begin{tabular}{l|rrrr}
\textbf{Source OSM (SM)} & \textbf{URLs} & \textbf{Tweets} & \textbf{Twitter Users} &  \textbf{SM Users} \cr \hline
\textbf{YouTube} & 34,306 & 141,118 & 88,231 & 26,026 \cr
\textbf{Foursquare} & 14,896 & 23,252 & 14,401 & - \cr
\textbf{Flickr} & 1,719 & 2,560 & 1,419 & 711 \cr
\textbf{Baseline} & - & 29,038,497 & 3,511,044 & - \cr
\end{tabular} \caption{\bf{Descriptive statistics of the datasets. In our dataset, YouTube is more popular on Twitter than Foursquare and Flickr. } \label{tab:datasets}}
\end{table}

In order to verify the representativeness of our datasets, we repeated the analysis using keywords related to another popular event in 2010 on Twitter -- the Brazilian Presidential Election.~\footnote{Dilma Rouseff, elected president of Brazil, was the second most cited person on Twitter in 2010.} We monitored this event, especially during the candidate's campaign, which started on July $6^{th}$ and ended on October $31^{st}$, the final election day. To monitor this event we used a set of 30 keywords (e.g. dilma, serra and marinasilva) related to the candidates and their political parties. Due to space constraints, we present results only for FWC datasets, but most of our conclusions hold for the Brazilian Presidential Election datasets as well.

%\vspace{-15pt}
%\input{results} % results
\section{Results} \label{sec:results}

\noindent In this section, we investigate three key questions about Cross-Pollination -- (1) What are the characteristics of Cross-Pollination? (2) Does Cross-Pollination across OSM services help to increase the audience reached by the information diffused? (3) What is the relationship between the OSM services involved and how does it affect the information diffusion process?

\subsection{Cross-Pollination characteristics}

\noindent Aiming at answering the first question, we analyze the characteristics of two groups -- temporal and topological. Each characteristic of Cross-Pollinated networks is studied separately and compared with diffusion OSM characteristics. We first discuss temporal characteristics of Cross-Pollination, and then we analyze topological characteristics.

\subsubsection{Sharing Activity}

\begin{figure}[!htb]  \begin{center}
\includegraphics[width=.49\textwidth] {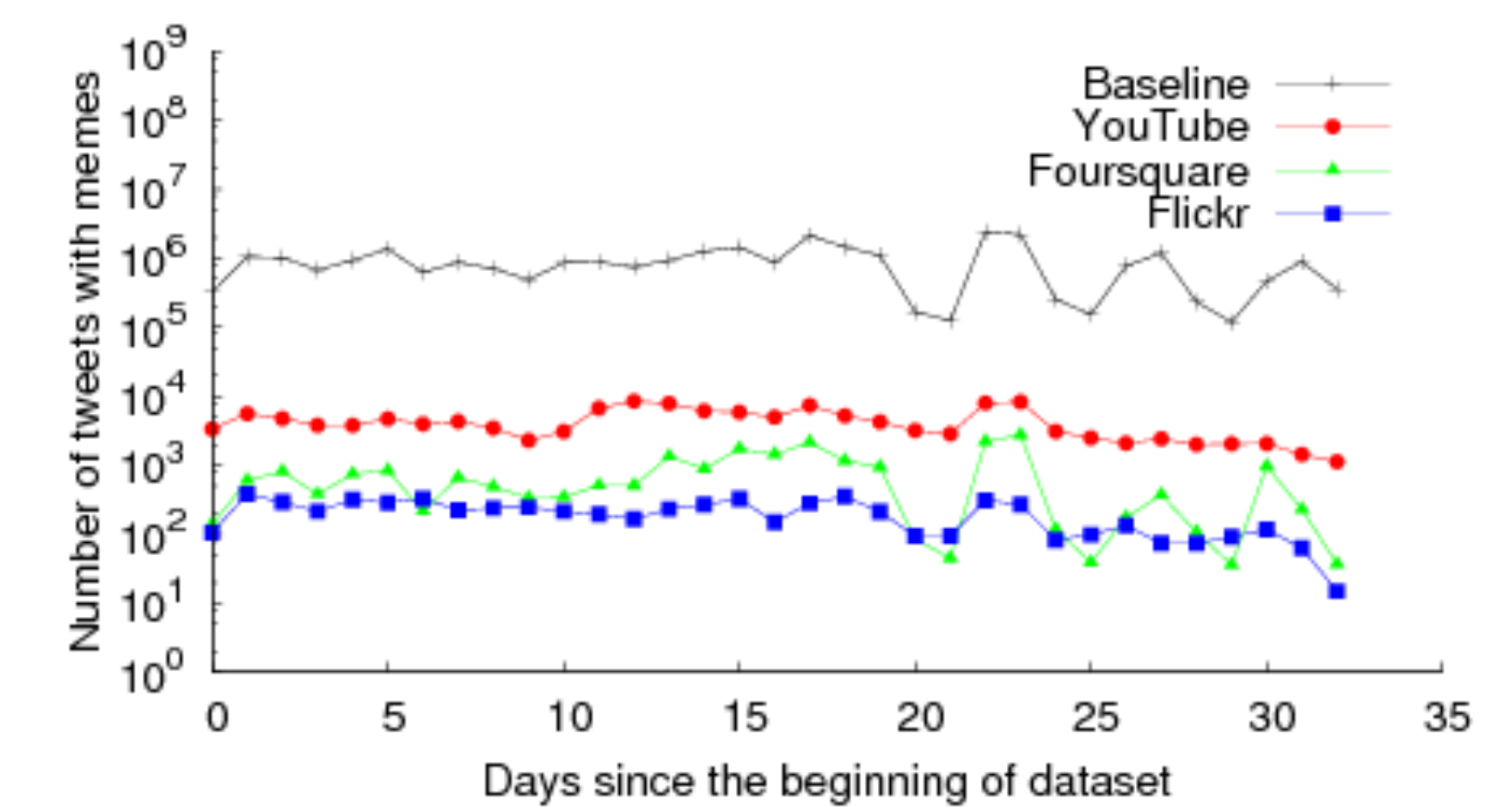}
\end{center}
\caption{\bf{Foreign and local meme creation over time. Show a relatively uniform creation activity during the complete period.}}
\label{fig:tweets_time}
\end{figure}

\noindent An important temporal characteristic
of Cross-Pollination is the volume of tweets generated by
foreign memes on a given day during a certain period of time.
Figure~\ref{fig:tweets_time} shows the total number of tweets with foreign memes created on each day during the FWC event. For comparison purposes, the figure also shows the total number of tweets with local memes created per day (using baseline dataset). We observe a similar trend during the whole period, for all datasets analyzed. The trend of volume of tweets created due to meme (both foreign and local) sharing is relatively uniform and similar during the whole period, with small peaks occurring on the same days. Hence, foreign meme sharing
activity follows local meme sharing activity, although absolute
numbers differ significantly (around $10^{3}$ YouTube foreign memes
on Twitter and $10^{6}$ local memes on baseline dataset).

\subsubsection{User participation}

\begin{figure}[!htb] \begin{center}
\includegraphics[width=.49\textwidth] {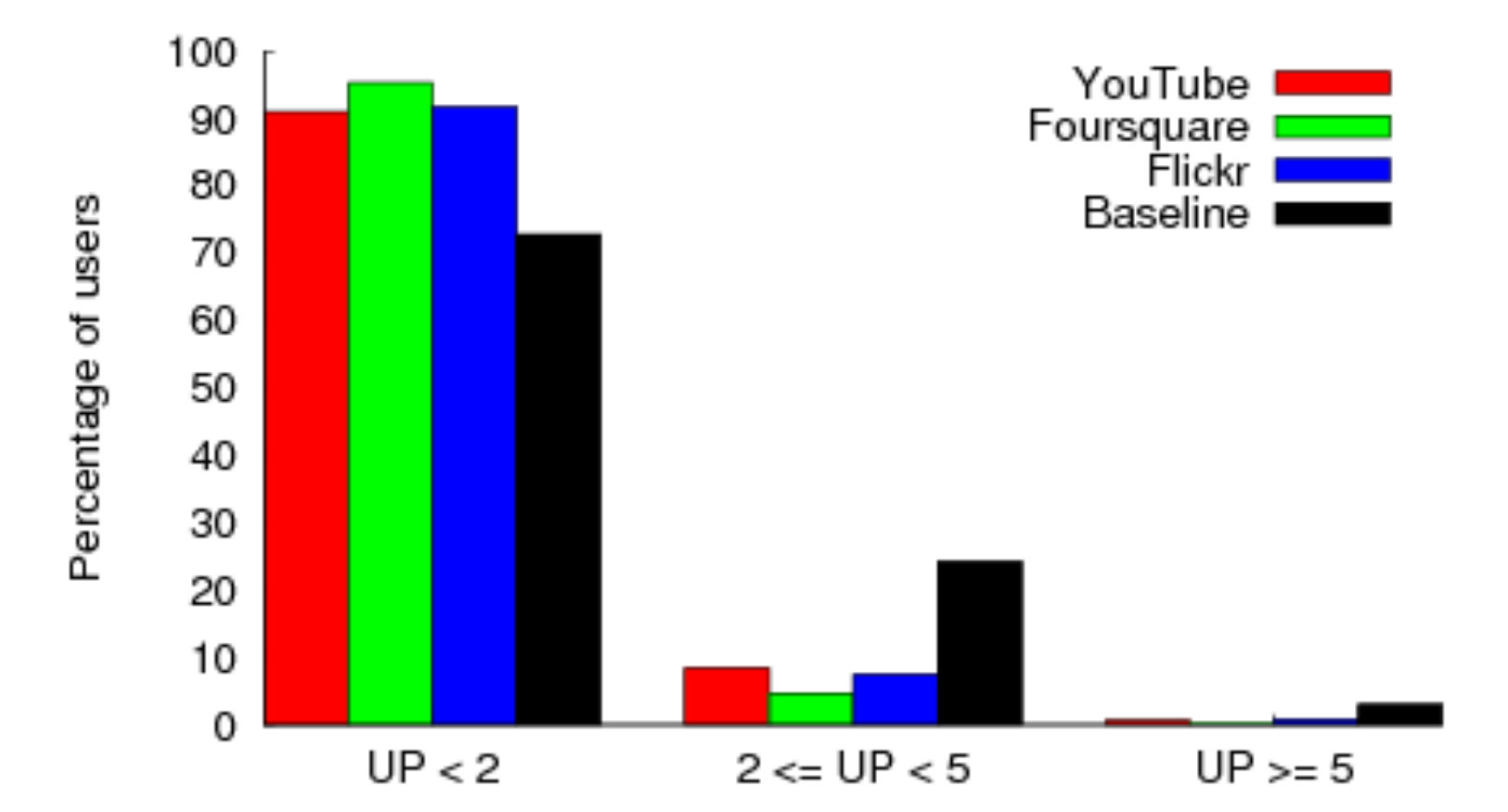}
\end{center}
\caption{\bf{User participation (UP) in meme creation activity. Most users have similar participation in meme creation for both local and foreign meme, as most users belong to the same bin.}}
\label{fig:user_participation}
\end{figure}

\noindent \emph{User Participation} (UP) in meme creation is defined as the average number of tweets with memes (foreign and local) created per day, for each user. Our goal with this metric is to check whether users contribute equally in the traffic generated by Cross-Pollination on the diffusion OSM. Towards this goal, we divided users into bins according to their UP, and then calculated the percentage of users in each bin (see Figure~\ref{fig:user_participation}). Users contribute equally for the traffic generated by Cross-Pollinated networks; vast majority of users (more than 90\%) are in the same bin, with less than 2 tweets with foreign memes created per day. Furthermore, Cross-Pollinated networks follow the diffusion OSM in this aspect, as the vast majority (more than 70\%) of users from the baseline dataset is on the same bin (UP < 2 tweets with local memes per day). It is much easier for a user to create a simple tweet with 140 characters than watch a video and create a tweet with its link. The same observation can be done for Flickr photos and Foursquare locations. We conjecture that this is the reason for the higher percentage of users with UP between 2 and 5 in the baseline dataset.

\subsubsection{Diffusion delay}

\begin{figure}[!htb]
\begin{center}
\includegraphics[width=.49\textwidth]{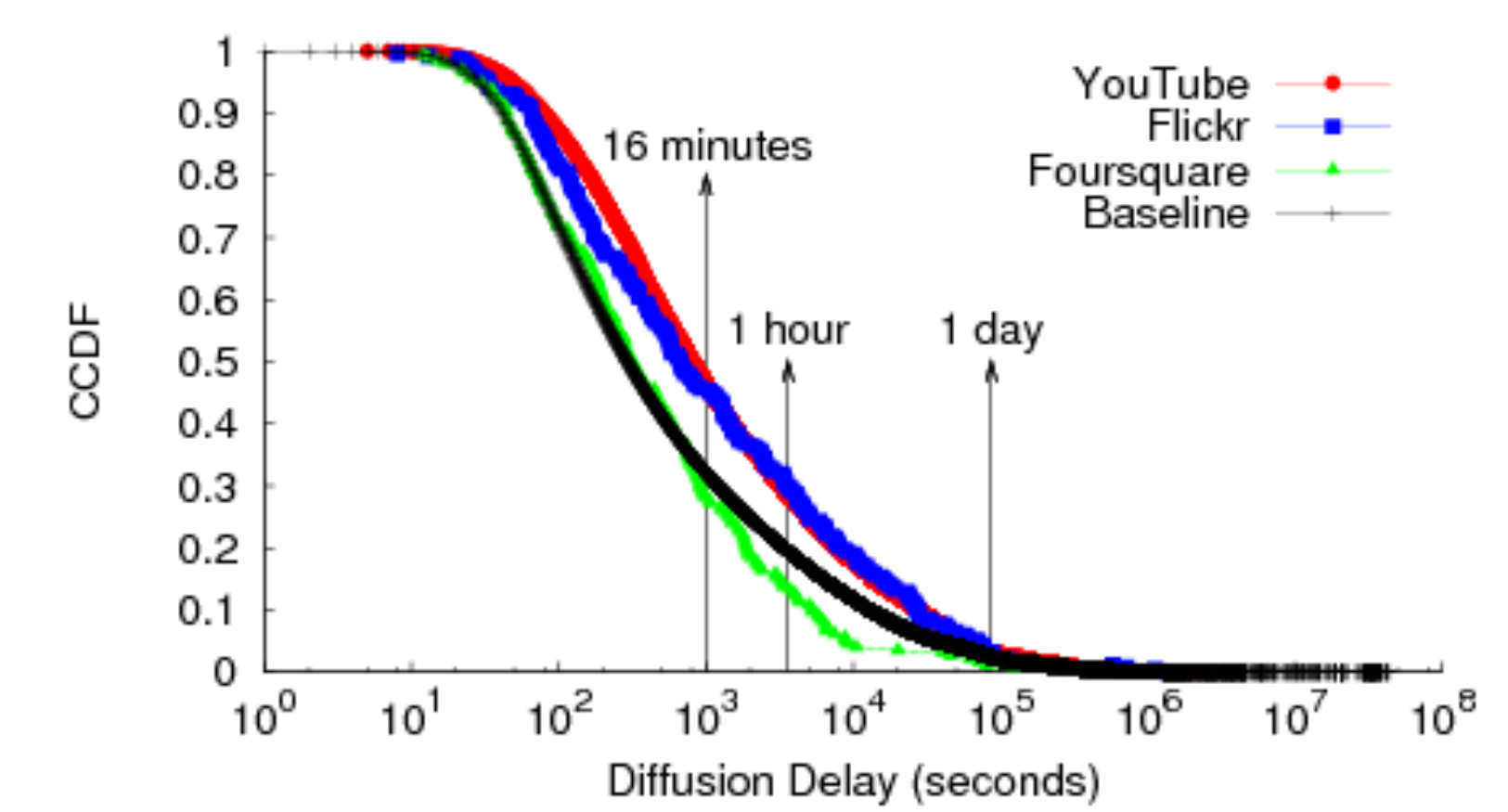}
\caption{\bf {Complementary cumulative distribution function of diffusion delay of memes. Show a similar distribution for both foreign and local memes. }}
\label{fig:diffusion_delay}
\end{center}
\end{figure}

\noindent We define \emph{diffusion delay} as the time between tweet meme getting created and being retweeted. Figure~\ref{fig:diffusion_delay} shows the complementary cumulative distribution function (CCDF) for the diffusion delay of the three Cross-Pollinated networks studied and the baseline dataset. On average, 75\% of the memes (both local and foreign) are retweeted in less than 1 hour, and 97\% are retweeted within a day. We note from the distributions that YouTube and Flickr memes tend to be retweeted with a slightly higher delay than Foursquare and local memes. For example, around 50\% of retweets from YouTube and Flickr memes have a delay larger than 1,000 seconds (around 16 minutes), while 30\% of retweets from Foursquare and local memes have a delay larger than 1,000 seconds. Nature of the content is a reasonable explanation for this difference. A user can easily read and quickly respond a direct message (local meme), while a foreign meme becomes an indirect message as the user is expected to view the content of the URL before forwarding it. In this case, Foursquare memes are more similar to local memes because they are usually automatically posted messages which contain the name of the place from where the user ``checked in'' together with the URL of the location. ``I am at DCC, UFMG http://4sq.com/XyZw'' is an example of this kind of tweet.

In conclusion from these analyses, Cross-Pollinated networks follow temporal characteristics of the diffusion OSM (Twitter), as similar results were found for both foreign and local memes sharing activity, user participation, and diffusion delay. Next, we analyze topological characteristics of Cross-Pollinated networks.

\subsubsection{Diffusion cascades}

\noindent We now turn our focus to analyze topological characteristics of Cross-Pollinated networks. Diffusion occurs through \emph{originators} and \emph{spreaders}. Originators are users who posted a foreign meme on Twitter, and spreaders are users who forwarded (i.e., either replied or retweeted) that foreign meme posted by an originator. A diffusion cascade is defined as a directed connected graph $G(V,E)$, where nodes represent originators / spreaders and edges represents that a foreign meme tweeted by an originator is forwarded by a spreader. Direction of the edge represents the information diffusion from originator to spreader. Only 18\% YouTube videos, 10\% Flickr photos, and 2\% Foursquare locations are forwarded at least once. 

\begin{figure}[!htb]
  \begin{center}
  \includegraphics[scale=.5]{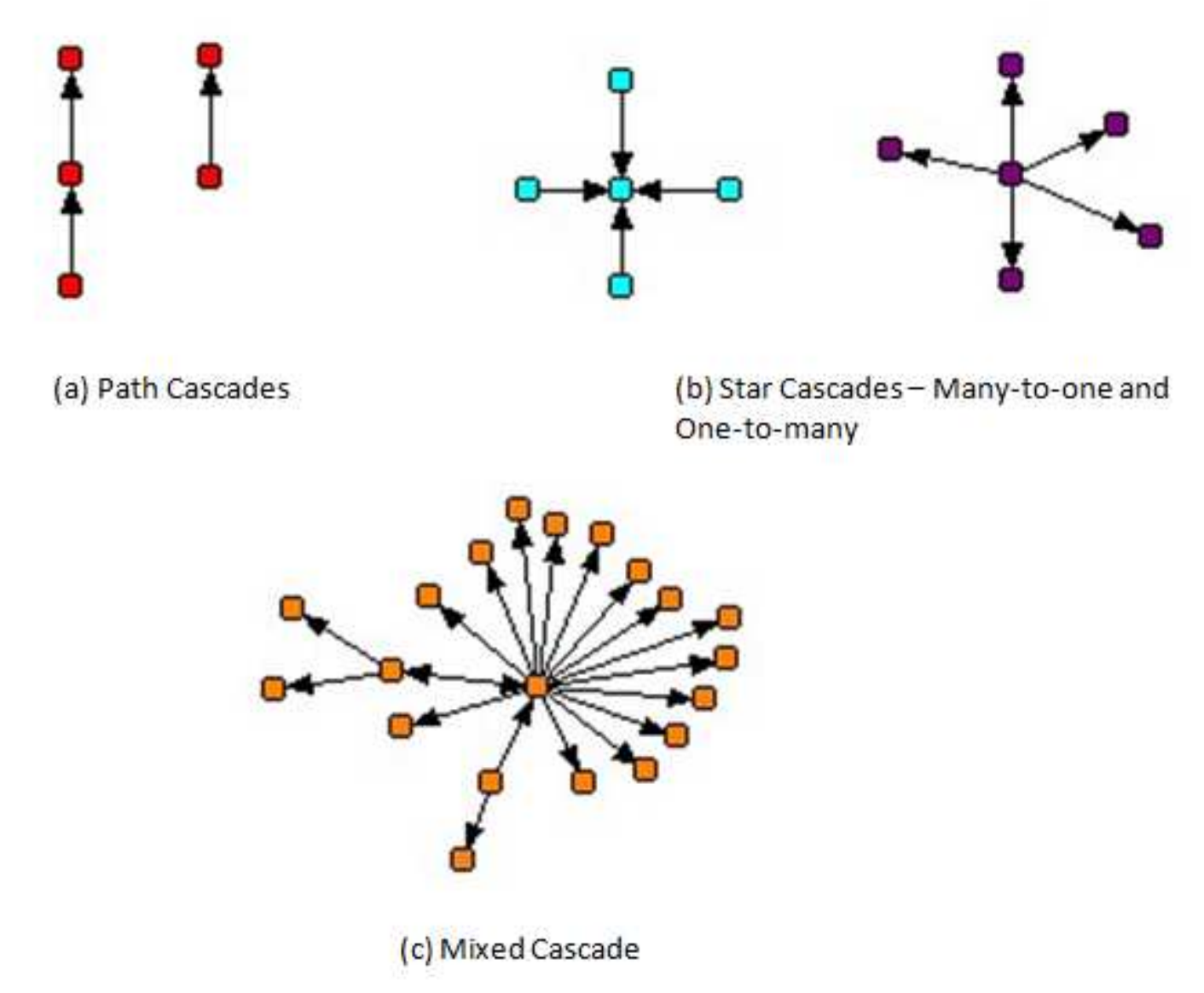}\\
  \caption{\bf{Diffusion cascades. Most of the foreign meme and local memes cascades are star-shaped followed by path and mixed directed connected cascades.}}
  \label{fig:topologies}
  \end{center}
\end{figure}

Foreign memes diffuse in cascades like star, path, and other connected cascades (see Figure~\ref{fig:topologies}), for the three Cross-Pollinated networks. Star cascades are formed when many users forward a single user's foreign meme, resulting in one-to-many diffusion of information. When one user forward many user's foreign meme, information diffuses from many to one user, resulting in many-to-one diffusion cascade. A path cascade is formed when a user forwards an already forwarded foreign meme, resulting in information diffusion from one user to another in a chain. A mixed connected cascade is formed when users involved in diffusion of foreign memes are associated with mixed actions of forwarding. Similar observations were found for the baseline dataset, where only 12\% of local memes are forwarded and diffused in path, star, and mixed cascades.

\begin{figure}[!htb]
  \begin{center}
  \includegraphics[width=.49\textwidth]{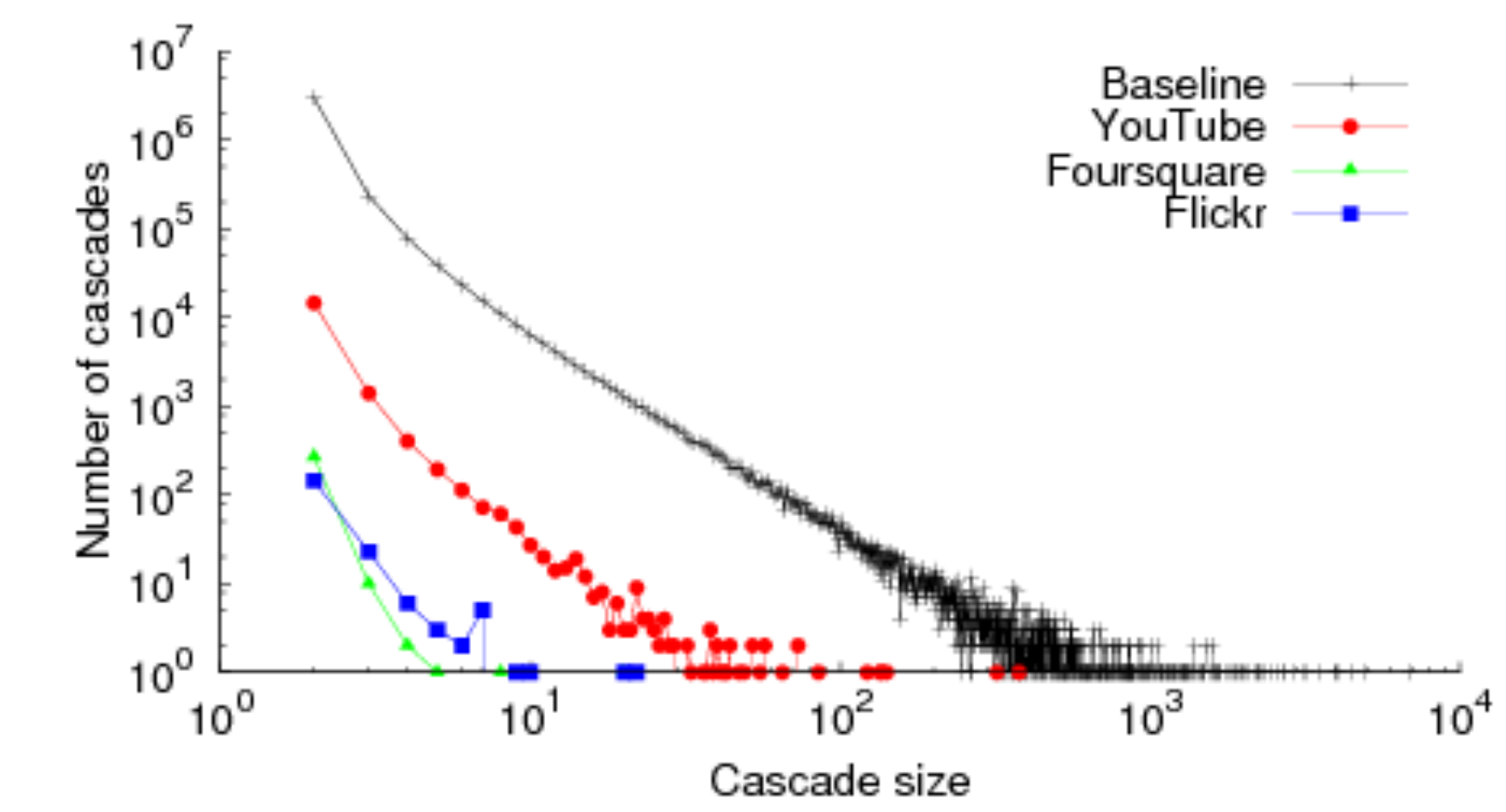}\\
  \caption{\bf{Distribution of cascades with cascade size. Shows a large number of foreign and local meme cascades with small size, while a small number of cascades with large size.}}
  \label{fig:component_distribution}
  \end{center}
\end{figure}

Figure~\ref{fig:component_distribution} shows distribution of number of cascades with cascade size. Out of the cascades formed by foreign memes diffused, most cascades are composed of only one originator and one spreader (i.e., of cascade size 2), resulting in diffusion of information only to one user (and user's followers), which we term as one level of diffusion. There are only few cascades which have large cascade size, reaching many users and users' followers. This implies that Cross-Pollination is not effective in diffusing the information, as most of the foreign memes remain unnoticed. For foreign memes that were diffused, the level of diffusion remains to only one user. Similar distribution can be seen for local memes. Number of cascades follow 90-10 Pareto distribution with 90\% cascades with size $\le$ 3 and 10\% cascades with size larger than 3, for both Cross-Pollinated network and baseline.

\subsubsection{Diffusion cascade statistics}

\noindent Table~\ref{tab:comparison_topology} shows a comparison of some graph metrics for diffusion cascades of the three CP networks in study and the baseline dataset. Twitter users are most attracted towards posting and forwarding YouTube videos than Flickr photos, Foursquare locations and local memes (average number of spreaders / meme is highest for YouTube -- 7.08, and highest average number of cascades / meme -- 2.53).  Even then, average diffusion cascade size for YouTube remains approximately 2.53, close to the other datasets in study.~\footnote{Spreaders can belong to different cascades for different foreign meme.} The reason is the large number of videos which get only one spreader, resulting in an average cascade size close to 2 (in accordance with Pareto principle followed by cascades distribution). Average in-degree and average out-degree for Cross-Pollinated networks are higher than baseline and close to each other. Hence, Cross-Pollinated networks behave similarly, irrespective of the type of foreign meme diffused.

\begin{table*}[!htb]
\centering
\scriptsize
\begin{tabular}{l|rrrrrrr}
\textbf{Source OSM} & \textbf{Degree} & \textbf{In-degree} & \textbf{Out-degree} & \textbf{Path Length} & \textbf{\# cascades / meme} & \textbf{Cascade Size} & \textbf{\# spreaders / meme} \cr
\hline 
\textbf{YouTube}     & 1.06 & 1.17 & 1.12 & 0.37 & 2.81 & 2.53 & 7.08 \cr \hline 
\textbf{Flickr}      & 1.11 & 1.06 & 1.48 & 0.43 & 1.11 & 2.69 & 2.97 \cr \hline 
\textbf{Foursquare}  & 1.03 & 1.09 & 1.06 & 0.48 & 1.02 & 2.13 & 2.18 \cr \hline
\textbf{Baseline}    & 1.07 & 0.53 & 0.53 & - & 1.00 & 2.78 & 2.78 \cr
\end{tabular} \caption{\bf{Diffusion cascade statistics for three Cross-Pollinated networks and baseline. All numbers are averages. Cascade characteristics are similar across Cross-Pollinated networks, implying its independence of the source OSM and are higher than baseline dataset.}} \label{tab:comparison_topology}
\end{table*}

In conclusion from these topological analyses, we observe that topological characteristics of Cross-Pollinated networks are independent of the source OSM and follow the diffusion OSM in most aspects.

\subsection{Relationship between the source and the diffusion OSM}

\noindent We turn our focus now to understand how the source and the diffusion OSM are related and how this relationship affects the meme diffusion process in a Cross-Pollinated network. Towards this goal, we intend to answer the following questions -- (1) Does popularity of a foreign meme in source OSM influence its popularity in diffusion OSM? (2) Does popularity of foreign meme in diffusion OSM affect its popularity on source OSM? (3) Does source and diffusion OSM share users, who are responsible for diffusing foreign memes from source to diffusion OSM? We start by studying and analyzing first question.

\subsubsection{Popularity influence of source on diffusion OSM}

\noindent We now analyze whether popular foreign memes on source OSM affects the foreign meme popularity on diffusion OSM. The popularity of a video on YouTube is measured by its view count. On Flickr, the popularity of a photo is also measured by its view count. The popularity of a location on Foursquare is measured by the number of ``check-ins.'' On Twitter, the popularity of a foreign meme is given by the number of tweets with it. We obtained the popularity ranking of foreign memes in both the source OSM and the diffusion OSM. In order to compare two rankings, we used the \textit{Kendall's Tau coefficient}~\cite{kendalltau}. The Kendall's tau coefficient is a measure of the rank correlation, i.e., the similarity of the orderings of the data when ranked by each of the quantities. It is given by the following equation:
\begin{equation}
\tau = \frac{\# of concordant pairs - \# of discordant pairs}{\frac{1}{2}.n.(n-1)}
\end{equation}
where $n$ is the total number of pairs. A concordant pair is a pair of elements in which the ranks agree, and a discordant pair is a pair of elements in which the ranks disagree. If the agreement between the two rankings is perfect, then the coefficient has value 1. If the disagreement between the two rankings is perfect, then the coefficient has value -1. If both rankings are independent, then the expected coefficient is approximately 0.

A $\tau$ of -0.0001 is observed on the Cross-Pollinated network between YouTube and Twitter, which demonstrates that the ranking of popularity of videos in both OSM services are independent. In other words, if a video is popular on YouTube does not mean that it will also be popular on Twitter, and vice-versa. Interestingly, we found same observation for Flickr and Foursquare datasets, as the $\tau$ is -0.0013 and -0.0001, respectively. Hence, popularity of foreign meme in source OSM does not influence its popularity in diffusion OSM.

\subsubsection{Popularity influence of diffusion on source OSM}

\noindent Towards answering the second question, Does popularity of foreign meme in diffusion OSM affect its popularity on source OSM?, we checked whether the diffusion of foreign memes in the diffusion OSM helps in increasing the traffic (popularity of memes) in the source OSM. Although our datasets do not have information of how many clicks each URL received on Twitter, some URL shorteners provide APIs with statistics of access to their links. One of the most popular services is \emph{http://bit.ly/}, which provides an API~\cite{bitlyapi}. We used the API to analyze how many clicks each meme shortened in a \emph{bit.ly} URL received from Twitter. In order to do this analysis, we collected data for all \emph{bit.ly} URLs of our dataset, and then checked how many clicks they received from the referrer \emph{twitter.com}. In total, we have 13,158 videos from YouTube dataset (38.4\% of total) and 1,719 photos from Flickr dataset (38.2\% of total) shortened with \emph{bit.ly}. We did not consider Foursquare dataset as we had only 37 \emph{bit.ly} URLs, which is not representative.~\footnote{Foursquare has its own URL shortener \emph{4sq.com}, which might be the reason for a small number of \emph{bit.ly} URLs in our dataset.}

\begin{figure}[!htb] \begin{center}
\includegraphics[width=.49\textwidth] {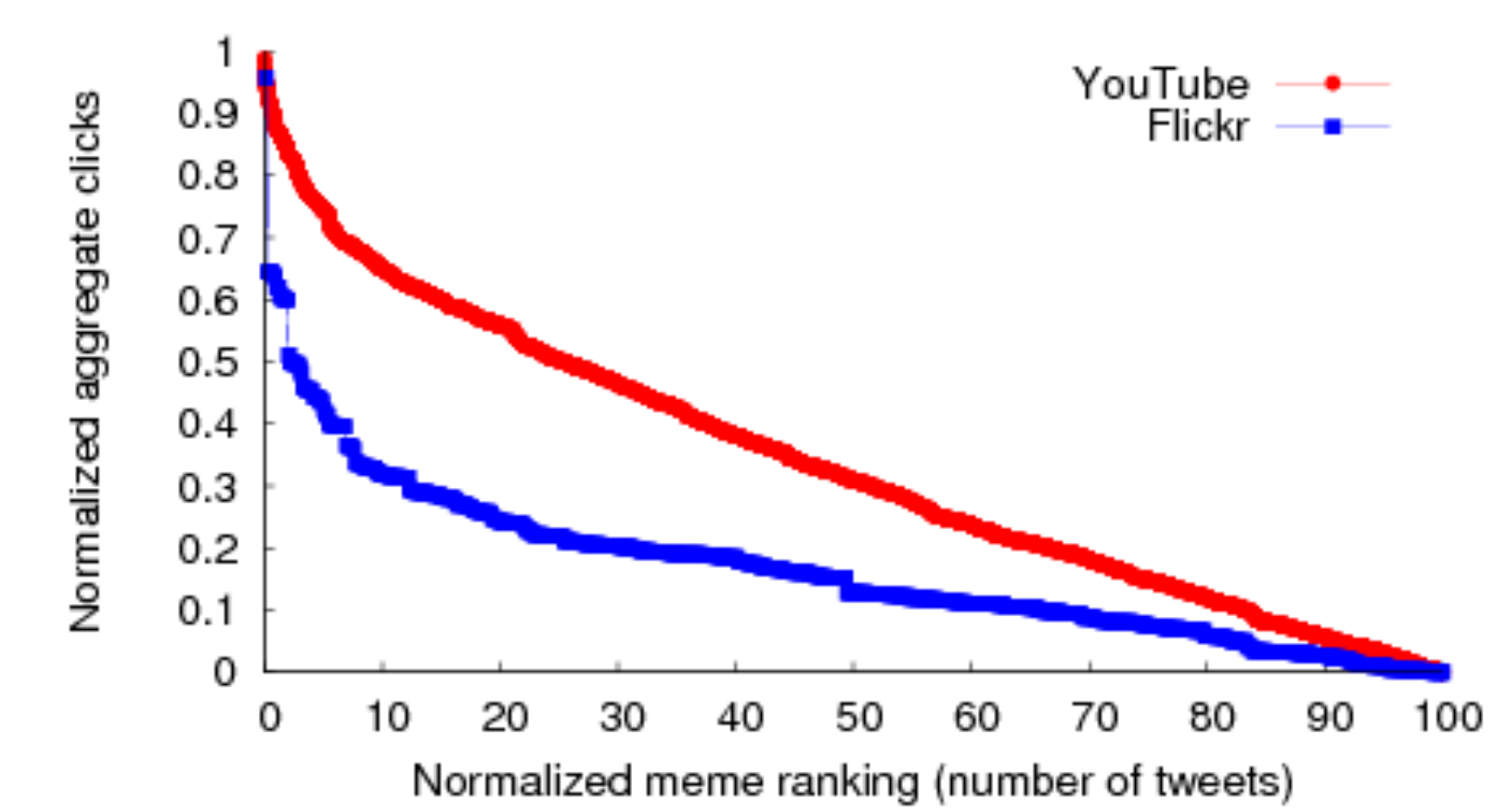}
\end{center}
\caption{\bf{Normalized aggregated clicks versus normalized foreign meme ranking. This implies that clicks for foreign memes from Twitter are not centered on popular foreign memes.} }
\label{fig:bitly_pareto_pop}
\end{figure}

We then analyze the fraction of views~\footnote{We consider that each click on the URL represents one view in the source OSM.} that each foreign meme (videos and photos) received from Twitter. We observe low fractions of views from Twitter, for many foreign memes tweeted. About 97\% of the videos received no more than 1\% of their views from Twitter, and almost 59\% of the photos received no more than 1\% of their views from Twitter. We then analyze if the most popular foreign  memes on Twitter (measured in number of tweets) are responsible for most part of the clicks to the source OSM. Figure~\ref{fig:bitly_pareto_pop} shows that the top 20\% most ``popular tweeted videos'' on Twitter are responsible for 56\% of the clicks, and that the top 20\% most ``popular tweeted photos'' on Twitter are responsible for 24\% of the clicks. This result implies that clicks for foreign memes from Twitter are not centered on popular foreign memes only, as most unpopular foreign memes are responsible for a considerable fraction of clicks. Our conjecture for this observation is the audience and influence of the user who created the meme on Twitter. If one highly influential user with a large audience (say 1,000,000 followers) publishes a video URL it can receive much more clicks that if several unpopular users publish it (say 100 users with 100 followers only).

Twitter does not seem to be effective in increasing the popularity of foreign meme on the source OSM. By including only clicks from \emph{bit.ly} URLs, we have a lower bound of the fraction of views that came from Twitter. There can be various other sources contributing to number of views for a foreign meme which we did not analyze here, but as \emph{bit.ly} URLs are popular in our datasets we do not expect different conclusions if we include more sources. 
Since source OSM and diffusion OSM are not related by popularity of memes, we next analyze if source OSM and diffusion OSM are related in some way through the users present on both the networks.

\subsubsection{Users presence on source and diffusion OSM}

\noindent Another interesting aspect of the relationship between the source OSM and the diffusion OSM is the presence of users on both networks. In order to estimate user presence, we checked, for all Twitter users who shared foreign memes, if their usernames exist on YouTube and Flickr. We also checked, for all YouTube and Flickr users who created videos and photos, respectively, if their usernames exist on Twitter. We refer to these users with presence in both the source OSM and the diffusion OSM as \textit{carriers}, as they might have brought foreign memes from the source OSM to the diffusion OSM. Table~\ref{tab:bridges} shows the number of Twitter users on each Cross-Pollinated network, the number of creators of foreign memes in the source OSM, and the number of carriers in each case (i.e., Twitter users and source OSM users). We did not include Foursquare in this analysis because a location is not associated to an owner. One limitation of this analysis is that a real person might use different usernames on different OSM services, but we believe that most part tend to use the same username as it will be easier to be found by friends. Another limitation is that a Twitter user might have access to the content of a source OSM without having an account on it.

Around 65\% of the Twitter users who shared YouTube videos have an account on YouTube; while only 10\% of the Twitter users who shared Flickr photos have an account on Flickr (see Table~\ref{tab:bridges}). A feasible impact of this difference is the higher popularity of YouTube videos shared on Twitter, but we let for future work to confirm this observation. Interestingly, only 4.2\% of YouTube users, creators of videos, have an account on Twitter, while 20.1\% of owners of photos on Flickr have an account on Twitter as well. On manual inspection, we found that most YouTube videos are of general interest, like comedies and music clips, while most part of Flickr photos are of personal interest. So, theoretically, YouTube videos attract interest of a higher number of users, who watch and share them on Twitter most frequently than Flickr photos, which are mostly shared by their own creators to a limited number of their friends.

\begin{table}[!htb]
\centering
\scriptsize
\begin{tabular}{l|rr|rr}
\textbf{Source OSM} & \textbf{Users} & \textbf{Carriers} & \textbf{Twitter users} & \textbf{Twitter Carriers} \cr \hline
\textbf{YouTube} & 28,721 & 1,207 (4.2\%) & 88,231 & 57,620 (65.3\%) \cr
\textbf{Flickr} & 711 & 143 (20.1\%) & 4,028 & 403 (10.0\%) \cr
\end{tabular} \caption{\bf{Statistics about carriers of information in Cross-Pollinated networks. Note that we have two types of carries -- (1) creators of content in source OSM; and (2) users who shared a URL with foreign memes on Twitter.}}
\label{tab:bridges}
\end{table}

Another interesting question about user presence is whether creators of contents in a source OSM make use of Cross-Pollination as an attempt to increase the traffic to their contents. For example, users might upload a video to YouTube and share a tweet with its link to announce to friends about the new video. We note that only 0.7\% of YouTube videos and only 7.4\% of Flickr photos were first tweeted by creators of contents in the source OSM, which implies that the audience (users who receive and watch videos and photos) play an important role in carrying the information from the source OSM to the diffusion OSM.

From these analyses, we learned that the relationship between the source and the diffusion OSM is weak and does not affect significantly the information diffusion process in Cross-Pollinated networks. The popularity of a foreign meme in the source OSM does not imply its popularity in the diffusion OSM, and vice-versa. Furthermore, the diffusion OSM (Twitter) benefits more from Cross-Pollinated networks than the source OSM, as a significant volume of traffic is created daily, involving high user participation on Twitter, but a small traffic is generated backwards to the source OSM.

\section{Related work} \label{sec:relatedwork}

\noindent In recent years, researchers contributed significantly in understanding the behavior of OSM. Blogs~\cite{Leskovec07},~YouTube~\cite{imc2007cha}, Facebook~\cite{Nazir_1}, and Digg~\cite{Szabo} are some of the OSM services which have been extensively studied. Blogging and micro-blogging networks are shown to have temporal and topological patterns which largely exhibit power law behavior~\cite{Leskovec07,outtwitting,kwak@www2010}. Cha~\textit{et al.} analyzed the blogging network structure and information diffusion patterns within the network. They also studied YouTube video diffusion through blogs but they did not explicitly analyze the diffusion network (Blogosphere)~\cite{cha_icwsm09}. Researchers have studied information diffusion process on Facebook through News Feeds but did not distinguish between types of News Feeds (e.g. photo or URL)~\cite{sun@icwsm2009}. Cha~\textit{et al.} provided an in-depth study of YouTube and other similar User Generated Content (UGC) systems and showed that UGC systems follow power law with truncated tails~\cite{imc2007cha}.

Twitter has recently gained attention of researchers due to its increased popularity~\cite{TwitterGrowth}. Krishnamurthy~\textit{et al.} presented a detailed characterization of Twitter~\cite{krishna@wosn08-1}, and Choudhury~\textit{et al.} analyzed how user similarities (homophily) along various other attributes can affect the information diffusion process~\cite{choudhury:birds-of-a-feather:-does-:2010:frzrm}. Both studies provided analysis of Twitter and information flow of foreign meme. Most of the studies on information diffusion in OSM are limited only to diffusion of local meme. Our research intends to fill this gap as we focus mainly on foreign meme diffusion in OSM.

Diffusion of information in OSM attracts a lot of attention from the research community. Liben-Nowell and Kleinberg~\cite{kleinberg-2008-ichain} reconstructed the propagation of massively circulated Internet chain letters and showed that their diffusion proceeds in a narrow but very deep-like pattern. Broxton~\textit{et al.} analyzed the diffusion of viral video popularity in social media, but focused only on how the popularity of a video varies with its introduction in social media. They concluded that viral videos gain popularity faster on OSM than through any other referring source or itself (e.g., search engines, YouTube, etc.), and that viral video popularity on Twitter is at a higher rate than in any other OSM website~\cite{broxton-2010}. With this background, we choose Twitter as a diffusion network and intend to study how YouTube, Flickr and Foursquare objects diffuse on Twitter, how such diffusion is supported and how the diffusion network changes. 

Some research work aim at modeling information diffusion in OSM. Gruhl \textit{et al.}~\cite{gruhl-www04} used the theory of infectious diseases to model the flow of information diffusion in the blogosphere, based on the use of keywords in blog posts. Adar and Adamic~\cite{adar-2005-tracking} further extended the idea of applying epidemiological models to describe the information flow. They relied on the explicit use of URLs between blogs to track the flow of information. We let the task of modeling information diffusion across OSM services for future work. 

\section{Discussion} \label{sec:conclusion}

\noindent In this paper we studied some properties of Cross-Pollinated networks. We believe that understanding these properties can help OSM service providers to improve or introduce new effective ways of sharing information across OSM services; to comprehend user involvement in information diffusion across OSM services; and to help users to chose a diffusion OSM in which they should share information, in an attempt to make it spread fast and effectively. Understanding Cross-Pollination also enhances an understanding of evolving information diffusion process across OSM services, which can be used for business perspectives.

Using datasets related to three popular OSM services (YouTube, Flickr and Foursquare), we demonstrated that Cross-Pollinated networks follow temporal and topological properties of the diffusion OSM (Twitter). We also showed that Cross-Pollinated networks are important to the diffusion OSM, involving a significant user participation in content creation every day. Another finding of our work is that the popularity of a foreign meme in the source OSM does not seem to affect its popularity in the diffusion OSM. We believe that our findings are helpful for marketers to explore the rich environment for advertisement purpose, and for social media providers to improve their systems and develop tools to facilitate the information exchange across networks. For example, Facebook already provides an interface which enable users a quick way to share photos, videos, and links with friends. New interfaces and applications to facilitate Cross-Pollination of information can be developed by other OSM service providers.

To the best of our knowledge, this work is the first step towards studying an emerging phenomenon in social media environments, with several future opportunities for researchers. We plan to provide a generalization of the Cross-Pollination and develop a formal model for the diffusion process in a Cross-Pollinated network. We also plan to extend some of our analysis. Firstly, we analyzed the Cross-Pollination considering the diffusion only in Twitter. Results involving different social media environments can present different characteristics. Secondly, we studied the relationship between the source OSM and the diffusion OSM only, but, for example, a video created on YouTube can be brought to Twitter by users who watched it on Facebook. Several other questions regarding the relationship between OSM services are let for future work. 

Although this research presented some interesting results about information diffusion across OSM services, it has some limitations. The datasets provide us a great opportunity to study the dynamics of a Cross-Pollinated network, but we only have data of tweets that contain some keyword used in the search. In order to reduce the impact of this limitation, a wide range of keywords related to the event were used, in several languages. Another limitation is related to Twitter API service -- it works as a stream, and sometimes it could be out of service, which made us lose some tweets. Fortunately, we did not have this problem for long periods. Identifying if the tweet is related with the event is another limitation, which was addressed by the research project from which we received the data.
\section{Acknowledgements} \label{sec:acknwl}

\noindent This work was supported by the Indo-Brazil Science Council. The authors would like to thank everyone who helped in collecting the data. The authors acknowledge all members of PreCog research group at IIIT-Delhi for their support.

{
\bibliographystyle{IEEEtran}
\bibliography{references}
}

\end{document}